\documentclass{amsart}
\usepackage{amssymb}
\newtheorem{theorem}{Theorem}[section]
\newtheorem{lemma}[theorem]{Lemma}
\newtheorem{proposition}[theorem]{Proposition}
\theoremstyle{definition}
\newtheorem{definition}[theorem]{Definition}
\newtheorem{remark}[theorem]{Remark}
\numberwithin{equation}{section}
\newcommand{\abs}[1]{\lvert#1\rvert}
\renewcommand{\l}{\mathcal{L}}
\def\sa{\mathcal{A}}
\def\h{\mathcal H}
\def\aut{\mathsf{Aut}}
\def\unit{\mathsf{U}}
\def\pro{\mathsf{P}}
\def\id{\mathsf{id}}
\def\p{\perp}
\def\t{\times}
\def\ot{\otimes}
\def\oto{\, \overline{\otimes}\, }
\def\aerts{\, {\land\kern-.85em\bigcirc}\,}
\begin{document}
\title{A characterization of the Aerts product of Hilbertian lattices}
\author{Boris Ischi}
\address{Boris Ischi, Laboratoire de
Physique des Solides, Universit\'e Paris-Sud, B\^atiment 510,
91405 Orsay, France}
\email{ischi@kalymnos.unige.ch}
\thanks{Supported by the Swiss National Science Foundation.}
\begin{abstract}
Let $\h_1$ and $\h_2$ be complex Hilbert spaces, $\l_1=\pro(\h_1)$
and $\l_2=\pro(\h_2)$ the lattices of closed subspaces, and let
$\l$ be a complete atomistic lattice. We prove under some weak
assumptions relating $\l_i$ and $\l$, that if $\l$ admits an
orthocomplementation, then $\l$ is isomorphic to the separated
product of $\l_1$ and $\l_2$ defined by Aerts. Our assumptions are
minimal requirements for $\l$ to describe the experimental
propositions concerning a compound system consisting of so called
{\it separated} quantum systems. The proof does not require any
assumption on the orthocomplementation of $\l$.
\end{abstract}
\maketitle
\section{Introduction}

In their 1936's founding paper on quantum logic, Birkhoff and von
Neumann postulated that the lattice describing the experimental
propositions concerning a quantum system is orthocomplemented (see
\cite{Birkhoff/Neumann:1936}, \S 9). We prove that this postulate
forces the lattice $\l_{sep}$ describing a compound system
consisting of so called {\it separated} quantum systems to be
isomorphic to the separated product defined by Aerts in
\cite{Aerts:1982}.

By separated we mean two systems (electrons, atoms or whatever)
prepared in different ``rooms'' of the lab, and before any
interaction take place. Recall that the state of a two-body system
$S$ can be either entangled or a product state. Any non-product
state violates a Bell inequality \cite{Gisin:1991}, hence for
separated systems as defined above, the state of $S$ is
necessarily a product. Whether the two systems are fermions or
bosons does not matter. Since they are prepared independently and
do not interact, they are distinguishable and not correlated.

It is important to note that our result does not require any
assumption on the orthocomplementation of $\l_{sep}$. Instead,
following Piron \cite{Piron:handbook} and Aerts \cite{Aerts:1982},
we assume $\l_{sep}$ to be complete and atomistic. Moreover, we
need some assumptions relating $\l_i$ and $\l_{sep}$ that
translate the fact that $\l_{sep}$ describes a compound system.
Such minimal conditions have been settle and studied first by
Aerts and Daubechies (see \cite{Aerts/Daubechies:1979}, \S 2), and
later by Pulmannov\'a \cite{Pulmannova:1983,Pulmannova:1985} and
Watanabe \cite{Watanabe:2003,Watanabe:2004}. We will see in
Section \ref{SectionComparison} that our assumptions are much
weaker than those of previous works. In Section
\ref{SectionSeparatedProduct} we recall the definition of the
separated product. In Section \ref{SectionS-TensorProducts} we
introduce our assumptions by defining what we call
$\mathsf{S}-$products. The main result is proved in Section
\ref{SectionOrthocomplementation}, whereas an important
preliminary result concerning automorphisms is established in
Section \ref{SectionAutomorphims}.
\section{The separated product}\label{SectionSeparatedProduct}

For terminology concerning lattice theory, we refer to
\cite{Maeda/Maeda:handbook}. We adopt the following notations. If
$\l$ is a complete atomistic lattice, and $a$ an element of $\l$,
then $\sa(a)$ denotes the set of atoms under $a$, and $\sa(\l)$
denotes the set of atoms of $\l$.  If $\l$ is moreover
orthocomplemented, then we denote the orthocomplementation by
$a\mapsto a^\p$. For atoms, we write $p\p q$ if $p\leq q^\p$.
Finally, the top and bottom elements are denoted by $1$ and $0$
respectively.

\begin{definition}[D. Aerts, \cite{Aerts:1982}]\label{DefinitionSeparatedProduct}
Let $\l_1$ and $\l_2$ be complete atomistic orthocomplemented
lattices. On $\sa(\l_1)\t\sa(\l_2)$ define the following binary
relation: $p\# q$ if and only if $p_1\p_1 q_1$ or $p_2\p_2 q_2$.
Then,
\[\l_1\aerts\l_2:=\{ R\subseteq \sa(\l_1)\t\sa(\l_2)\,;\,\
R^{\#\#}=R\}\, .\]
\end{definition}

\begin{remark}Obviously, $\#$ is symmetric, anti-reflexive and {\it
separating} ({\it i.e.} for all $p\ne q$, there is $r$ with $p\#
r$ and $q\,\backslash\hspace{-2.3mm}\# r$), therefore
$\l_1\aerts\l_2$ is a complete atomistic lattice, the mapping
$a\mapsto a^\#$ of $\l_1\aerts\l_2$ into itself, where
$a^\#=\{p\in\sa(\l_1)\t\sa(\l_2)\,;\,p\# q,\,\forall q\in a\}$, is
an orthocomplementation, and atoms of $\l_1\aerts\l_2$ are
singletons of $\sa(\l_1)\t\sa(\l_2)$. Moreover, coatoms are given
by
\[\{(p_1, p_2)\}^\#=\sa(p_1^{\p_1})\t \sa(\l_2)\ \cup\ \sa(\l_1)\t
\sa(p_2^{\p_2})\, .\]
Hence, it is an easy exercise to prove that
\begin{equation}\l_1\aerts\l_2=\{\cap\omega\,;\,\omega
\subseteq\{\sa(a_1)\t\sa(\l_2)
\cup\sa(\l_1)\t\sa(a_2)\,;\, (a_1,a_2)\in\l_1\t\l_2\}\}\,
.\label{EquationEquivalentDefinitionSeparatedProduct}\end{equation}
For complete atomistic lattices $\l_1$ and $\l_2$, we define
$\l_1\aerts\l_2$ by
(\ref{EquationEquivalentDefinitionSeparatedProduct}).
\end{remark}
\section{$\mathsf{S}-$products}\label{SectionS-TensorProducts}

For our main result (Theorem \ref{TheoremTheTheorem}), we need to
make some hypotheses on $\l_1$ and $\l_2$, which are true if
$\l_1=\pro(\h_1)$ and $\l_2=\pro(\h_2)$, with $\h_1$ and $\h_2$
complex Hilbert spaces. However, we consider a more general
setting in order to point out exactly the assumptions on $\l_1$
and $\l_2$ needed for the proof.

Let $\l$ be a complete atomistic lattice. We write $\aut(\l)$ for
the group of automorphisms of $\l$. We say that $\l$ is {\it
transitive} if the action of $\aut(\l)$ on $\sa(\l)$ is
transitive. We denote by $2$ the lattice with two elements. If
$\h$ is a complex Hilbert space, then $\pro(\h)$ denotes the
lattice of closed subspaces of $\h$ and $\unit(\h)$ stands for the
group of automorphisms of $\pro(\h)$ induced by unitary maps.

\begin{remark} Let $\h$ be a complex Hilbert space. Then
$\unit(\h)$ acts transitively on $\sa(\pro(\h))$.\end{remark}

\begin{definition}\label{DefinitionConnected}
A complete atomistic lattice $\l$ is {\it weakly connected} if
$\l\ne 2$ and if there is a {\it connected covering} of $\sa(\l)$,
that is a family of subsets
$\{A^\gamma\subseteq\sa(\l)\,;\,\gamma\in\sigma\}$ such that
\begin{enumerate}
\item[(1)] $\sa(\l)=\cup \{A^\gamma\,;\,\gamma\in \sigma\}$ and
$\abs{ A^\gamma}\geq 2$ for all $\gamma\in\sigma$,
\item[(2)] for all $\gamma\in\sigma$ and for all $p\ne q\in
A^\gamma$, $p\vee q$ contains a third atom,
\item[(3)] for all $p,\,q\in\sa(\l)$, there is a finite set
$\{\gamma_1,\cdots,\gamma_n\}\subseteq\sigma$ such that $p\in
A^{\gamma_1}$ and $q\in A^{\gamma_n}$, and such that
$\abs{A^{\gamma_i}\cap A^{\gamma_{i+1}}} \geq 2$ for all $1\leq
i\leq n-1$.
\end{enumerate}
\end{definition}

\begin{remark} Note that in part 2 of Definition
\ref{DefinitionConnected}, it is not required that the third atom
under $p\vee q$ is in $A^\gamma$. Let $\l$ be a complete atomistic
lattice. If $\l$ is weakly connected, then $\l$ is irreducible
(see \cite{Maeda/Maeda:handbook}, Theorem 4.13). On the other
hand, any complete atomistic orthocomplemented irreducible lattice
$\l\ne 2$ with the covering property (for instance $\pro(\h)$ with
$\h$ a complex Hilbert space with $\mathrm{dim}(\h)\geq 2$) is
weakly connected. Indeed, in that case, for any two atoms $p$ and
$q$ of $\l$, $p\vee q$ contains a third atom, hence $\{\sa(\l)\}$
is a connected covering.
\end{remark}

\begin{definition}
Let $\l_1$, $\l_2$ and $\l$ be complete atomistic lattices and let
$h_1:\l_1\rightarrow\l$ and $h_2:\l_2\rightarrow\l$ be injective
maps preserving all meets and joins. For $a_1\in\l_1$,
$a_2\in\l_2$, $A_1\subseteq\sa(\l_1)$ and $A_2\subseteq\sa(\l_2)$,
we define
\[\begin{split}
a_1\ot a_2&:=(h_1(a_1)\land h_2(a_2))\, ,\\
a_1\oto a_2&:=\{p_1\ot p_2\,;\,p_1\in\sa(a_1),\,
p_2\in\sa(a_2)\}\, ,\\
a_1\oto A_2&:=\{p_1\ot p_2\,;\,p_1\in\sa(a_1),\,
p_2\in A_2\}\, ,\\
A_1\oto A_2&:=\{p_1\ot p_2\,;\,p_1\in A_1,\, p_2\in A_2\}\, .
\end{split}\]
\end{definition}

\begin{remark} Since $h_i$ preserves arbitrary joins and
meets, $h_i$ also preserves $0$ and $1$. Therefore,
$h_1(a_1)=h_1(a_1)\land h_2(1)=a_1\ot 1$ and $h_2(a_2)=h_1(1)\land
h_2(a_2)=1\ot a_2$. Moreover, $0\ot a_2=0=a_1\ot 0$ for all
$a_i\in\l_i$.\end{remark}

\begin{definition}\label{DefinitionLatterallyConnected}
Let $\l_1,\,\l_2$ and $\l$ be complete atomistic lattices, with
$\l_1$ and $\l_2$ weakly connected, and let
$h_1:\l_1\rightarrow\l$ and $h_2:\l_2\rightarrow\l$ be injective
maps preserving all meets and joins. Suppose that $p_1\ot
p_2\in\sa(\l)$, for all $(p_1,p_2)\in\sa(\l_1)\t\sa(\l_2)$.

We say that $\l$ is {\it laterally connected} if there is a
connected covering $\{A_1^\gamma\}_{\gamma\in\sigma_1}$ of
$\sa(\l_1)$ and a connected covering
$\{A_2^\gamma\}_{\gamma\in\sigma_2}$ of $\sa(\l_2)$, such that for
all $\gamma_1\in\sigma_1$ and $\gamma_2\in\sigma_2$, and for all
$q_1\ot q_2,\, r_1\ot r_2\in A_1^{\gamma_1}\oto A_2^{\gamma_2}$
with $q_1\ne r_1$ and $q_2\ne r_2$, there is ($\exists$) $p=p_1\ot
p_2\in\sa(\l)$ such that both lateral joins $p_1\ot q_2\vee p_1\ot
r_2$ and $q_1\ot p_2\vee r_1\ot p_2$ contain a third atom.

In case $\l_1=\pro(\mathbb{C}^2)$ (respectively
$\l_2=\pro(\mathbb{C}^2)$), we require moreover that for any atom
$p_1\in\sa(\l_1)$ (respectively $p_2\in\sa(\l_2)$), there is
($\exists$) $q\in\sa(\l_2)$ (respectively $r\in\sa(\l_1)$) such
that $p_1\ot q\vee p_1^{\p_1}\ot q$ (respectively $r\ot p_2\vee
r\ot p_2^\p$) contains a third atom.
\end{definition}

\begin{definition}\label{DefinitionC-TensorProduct}
Let $\l_1$, $\l_2$ and $\l$ be complete atomistic lattices with
$\l_1$ and $\l_2$ weakly connected. Let $T_i\subseteq\aut(\l_i)$.
We write $\l\in\mathbf{C}_{_{T_1T_2}}(\l_1,\l_2)$ if
\begin{enumerate}
\item[(P0)] there exist two injective maps $h_i:\l_i\rightarrow\l$
preserving all meets and joins,
\item[(P1)] $p_1\ot p_2\in\sa(\l)$, $\forall p_i\in\sa(\l_i)$,
\item[(P2)] $p_1\ot p_2\leq a_1\ot 1\vee 1\ot a_2\Leftrightarrow
p_1\leq a_1$ or $p_2\leq a_2$, $\forall
p_i\in\sa(\l_i),\,a_i\in\l_i$,
\item[(P3)] $\l$ is laterally connected,
\item[(P4)] for all $u_i\in T_i$, there is $u\in\aut(\l)$ such
that $u(p_1\ot p_2)=u_1(p_1)\ot u_2(p_2)$, $\forall
p_i\in\sa(\l_i)$.
\end{enumerate}
We call $\l$ a {\it $\mathsf{S}_{_{T_1T_2}}-$product} if
$\l\in\mathbf{C}_{_{T_1T_2}}(\l_1,\l_2)$ and
\begin{enumerate}
\item[(P5)] $\sa(\l)=\{p_1\ot
p_2\,;\,(p_1,p_2)\in\sa(\l_1)\t\sa(\l_2)\}$.
\end{enumerate}
\end{definition}

\begin{remark} Let $\l$ be a $\mathsf{S}-$product of $\l_1$
and $\l_2$. If $\l_1=2$, then $\l\cong\l_2$, and if $\l_2=2$ ,
then $\l\cong\l_1$. By Axiom P5, the $u$ of Axiom P4 is
necessarily unique. We denote it by $u_1\ot u_2$. By Axioms P4 and
P5, if $T_1$ acts transitively on $\sa(\l_1)$ and $T_2$ acts
transitively on $\sa(\l_2)$, then $\l$ is transitive. Therefore,
if $T_1$ and $T_2$ contain the identity, then the $\exists$ in
Definition \ref{DefinitionLatterallyConnected} can be replaced by
$\forall$. Finally, note that Axiom P3 requires that only some
lateral joins of atoms contain a third atom.
\end{remark}

The proof of the following proposition is left as an exercise.

\begin{proposition} Let $\h_1$ and $\h_2$ be complex Hilbert
spaces. Then $\pro(\h_1\ot
\h_2)\in\mathbf{C}_{_{\unit(\h_1)\unit(\h_2)}}
(\pro(\h_1),\pro(\h_2))$.\end{proposition}

\begin{remark}\label{RemarkLatteralJoinsinAerts} Let $\l_1$ and
$\l_2$ be complete atomistic lattices.
Note that from
(\ref{EquationEquivalentDefinitionSeparatedProduct}) we find that
lateral joins of atoms in $\l_1\aerts\l_2$ are given by $\{(p_1,
p_2)\}\vee \{(p_1,q_2)\}=\{p_1\}\t\sa(p_2\vee q_2)$ and $\{(p_1,
p_2)\}\vee \{(q_1, p_2)\}=\sa(p_1\vee q_1)\t \{p_2\}$. Moreover,
let $\omega\subseteq\l_1$. Then, $\vee\{\sa(a)\t\sa(\l_2)\,;\,
a\in\omega\}=\sa(\vee\omega)\t\sa(\l_2)$.
\end{remark}

\begin{lemma}\label{LemmaAertss-TensorProduct} Let $\l_1$ and
$\l_2$ be complete atomistic weakly connected lattices. Then,
$\l_1\aerts\l_2$ is a $\mathsf{S}_{_{T_1T_2}}-$product of $\l_1$
and $\l_2$ with $T_i=\aut(\l_i)$.\end{lemma}

\begin{proof} Define $h_1:\l_1\rightarrow\l_1\aerts\l_2$ as
$h_1(a_1):=\sa(a_1)\t\sa(\l_2)$, and
$h_2\rightarrow\l_1\aerts\l_2$ as $h_2(a_2)=\sa(\l_1)\t\sa(a_2)$.
From the preceding remark, $h_1$ and $h_2$ obviously preserve
arbitrary meets and joins, and Axiom P3 holds. Moreover, by
definition (see
(\ref{EquationEquivalentDefinitionSeparatedProduct})), Axioms P5
and P2 hold.

Finally, let $u_1$ be an automorphism of $\l_1$ and $u_2$ an
automorphism of $\l_2$. Define a map $u$ on $\sa(\l_1)\t\sa(\l_2)$
as $u(p_1,p_2)=(u_1(p_1),u_2(p_2))$. Then, for all
$(a_1,a_2)\in\l_1\t\l_2$, we have that
\[u(\sa(a_1)\t\sa(\l_2)\cup\sa(\l_1)\t\sa(a_2))=\sa(u_1(a_1))
\t\sa(\l_2)\cup \sa(\l_1)\t\sa(u_2(a_2))\, .\]
Moreover, $u$ preserves arbitrary set-intersections, hence induces
an automorphism of $\l_1\aerts\l_2$.\end{proof}

\begin{remark}
Given two complete atomistic weakly connected lattices with the
covering property $\l_1$ and $\l_2$, there are many
$\mathsf{S}-$products of $\l_1$ and $\l_2$ different from
$\l_1\aerts\l_2$ \cite{Ischi:2004} (note that Axiom P3 in
\cite{Ischi:2004} is stronger than Axiom P3 here).
\end{remark}

\begin{lemma}\label{LemmaJoinins-Tensor} Let $\l_1,\,\l_2$ and
$\l$ be complete atomistic lattices. Suppose that $\l$ is a
$\mathsf{S}-$product of $\l_1$ and $\l_2$. Let $p_1\ot p_2$,
$q_1\ot q_2\in\sa(\l)$, $a_1\in\l_1$ and $a_2\in\l_2$. Suppose
that $p_1\ne q_1$, $p_2\ne q_2$ and that $p_1\leq a_1$ and
$p_2\leq a_2$. Then
\begin{enumerate}
\item $p_1\ot p_2\vee q_1\ot q_2$ contains no third atom,
\item $\sa(p_1\ot a_2\vee a_1\ot p_2)=p_1\oto a_2\cup a_1\oto
p_2$.
\end{enumerate}
\end{lemma}

\begin{proof}
(1) First,
\[p_1\ot p_2\vee q_1\ot q_2\leq (p_1\ot 1\vee 1\ot q_2)\land
(q_1\ot 1\vee 1\ot p_2)\, .\]
Now, from Axioms P2 and P5 we find that
\[\begin{split}\sa((p_1\ot 1\vee 1\ot q_2)\land (q_1\ot 1\vee 1\ot p_2))
&=(p_1\oto 1\cup 1\oto q_2)\cap(q_1\oto 1\cup 1\oto p_2)\\
&=\{p_1\ot p_2,q_1\ot q_2\}\, .\end{split}\]

(2) First,
\[p_1\ot a_2\vee a_1\ot p_2\leq a_1\ot 1\land 1\ot a_2\land(p_1\ot
1\vee 1\ot p_2)\, ,\]
and by Axioms P2 and P5,
\[\begin{split}\sa(a_1\ot 1\land 1\ot a_2\land(p_1\ot
1\vee 1\ot p_2))&=a_1\oto a_2\cap(p_1\oto 1\cup 1\oto p_2)\\
&=p_1\oto a_2\cup a_1\oto p_2\, .\end{split}\]
\end{proof}
\section{$\mathsf{S}-$products and separated quantum systems}\label{SectionComparison}

In this section we discuss and compare our Axioms listed in
Definition \ref{DefinitionC-TensorProduct} with those of previous
works.

Let $\l_1$, $\l_2$ and $\l$ be complete atomistic
orthocomplemented lattices. In
\cite{Aerts/Daubechies:1979,Pulmannova:1983,Pulmannova:1985,Watanabe:2003,Watanabe:2004}
it is required for $\l$ to describe a compound system that
\begin{enumerate}
\item[(l0)] $\l_1$, $\l_2$ and $\l$ are orthomodular,
\item[(p0)] there exists two injective maps
$h_i:\l_i\rightarrow\l$ preserving the orthocomplementation and
all meets and joins,
\item[(p1)] $p_1\ot p_2\in\sa(\l)$, $\forall p_i\in\sa(\l_i)$,
\item[(p2)] for all $(a_1,a_2)\in\l_1\t\l_2$, $h_1(a_1)$  and
$h_2(a_2)$ commute.
\end{enumerate}
Obviously, Axiom p1 is identical to Axiom P1 and Axiom p0 implies
axiom P0. On the other hand, from Axioms l0, p0, p1 and p2 follows
that $p_1\ot p_2\leq h_1(a_1)\vee h_2(a_2)$ if and only if
$p_1\leq a_1$ or $p_2\leq a_2$. [{\it Proof}\nobreak : The proof
is very similar to the proof of Lemma 1 in \cite{Pulmannova:1985}.
Let $S=\{h_1(p_1),h_2(p_2),h_1(a_1),h_2(a_2)\}$. From Axiom p2,
for all $T\subseteq S$ with $\abs{T}=3$, there is $a\in T$ such
that $a$ commutes with all $b\in T$. Therefore, since $\l$ is
orthomodular, the sublattice generated by $S$ is distributive
\cite{Greechie:77}. Suppose that $p_2\land a_2=0$. Then,
\[p_1\ot
p_2\leq (h_1(a_1)\vee h_2(a_2))\land h_2(p_2)=a_1\ot p_2\vee 1\ot
(a_2\land p_2)=a_1\ot p_2\, .\]

Let $u:\l_1\rightarrow \l$ be defined as $u(a)=a\ot p_2$. By Axiom
p1, $u(a)=0$ implies $a=0$. Moreover, since $\l$ is orthomodular,
the map $\,^*$ from the sublattice $[0,h_2(p_2)]$ into itself
defined as $x^*=x^\p\land h_2(p_2)$ is an orthocomplementation,
and $[0,h_2(p_2)]$ is orthomodular. Hence, from Axiom p0,
$u(a^{\p_1})=h_1(a)^*$. On the other hand, by Axiom p2,
$u(a)^*=(h_1(a_1)^\p\vee h_2(p_2)^\p)\land h_2(p_2)=h_1(a)^*$
\cite{Greechie:77}.

As a consequence, since $u(p_1)\leq u(a_1)$, we find that
$u(p_1)\land(u(p_1)\land u(a_1))^*=0$, therefore $p_1\land
(p_1\land a_1)^{\p_1}=0$; whence, since $\l_1$ is orthomodular,
$a\land p_1=(a_1\land p_1)\vee(p_1\land(p_1\land
a_1)^{\p_1})=p_1$, that is $p_1\leq a_1$. ]

As a consequence, Axioms l0, p0, p1 and p2 imply Axiom P2.
Therefore, from Axiom p0, we find that $p_1\ot p_2\leq (q_1\ot
q_2)^\p$ if and only if $p_1\p_1 q_1$ or $p_2\p_2 q_2$. Hence,
from Axioms l0, p0, p1, p2 and P5, we find that
$\l\cong\l_1\aerts\l_2$, which by Lemma
\ref{LemmaAertss-TensorProduct} is a $\mathsf{S}-$product of
$\l_1$ and $\l_2$ if $\l_1$ and $\l_2$ are weakly connected.

In \cite{Ischi:2004,Ischi:2002} we proved a similar result as
here. However, the proof in \cite{Ischi:2002} requires an axiom
relating the orthocomplementations of $\l_i$ and $\l$, whereas in
\cite{Ischi:2004} we used an axiom P3 stronger than Axiom P3 here.

We now make some comments about our axioms. Let $\l_1=\pro(\h_1)$
and $\l_2=\pro(\h_2)$ with $\h_1$ and $\h_2$ complex Hilbert
spaces, and let $\l$ be a complete atomistic lattice describing
the experimental propositions concerning a compound system $S$
consisting of two separated quantum systems $S_1$ and $S_2$,
described by $\l_1$ and $\l_2$ respectively.

As mentioned in the introduction, since $S_1$ and $S_2$ are
separated, Axiom P5 holds. On the other hand, Axiom P2 can be
justified easily (see \cite{Aerts:1982} or \cite{Ischi:2002} for
details), and Axioms P0 and P4 with $T_i=\unit(\h_i)$ are indeed
very natural.

Axiom P3 is more delicate. At a first glance, it may appear
technical. However, there is a simple physical reason why $\l$
should be laterally connected. Indeed, it is natural to assume
that there is a map $\omega:\sa(\l)\t\l\rightarrow [0,1]$ which
satisfies at least the two following hypotheses:
\begin{enumerate}
\item[(A1)] $\omega(p,a)=1\Leftrightarrow p\leq a$,
\item[(A2)] $\omega(p_1\ot p_2, a_1\ot
a_2)=g(p_1,a_1)g(p_2,a_2)$,\end{enumerate}
with $g(p,a)=\Vert P_a(v)\Vert^2$, where $P_a$ denotes the
projector on $a$, and $v$ is any normalized vector in $p$. Hence,
for all atoms $p_1,\, p\in\sa(\l_1)$ and $p_2,\,r,\,
s\in\sa(\l_2)$, such that $r\p_2 s$, we have
\begin{equation}\begin{split}\omega(p_1\ot p_2,p\ot (r\vee
s))&=g(p_1,p)g(p_2,r\vee s)\\
&=g(p_1,p)(g(p_2,r)+g(p_2,s))\\
&=\omega(p_1\ot p_2,p\ot r)+\omega(p_1\ot p_2,p\ot s)\,
.\end{split}\label{EquationPropensities}\end{equation}

On the other hand, for any two orthogonal atoms $r$ and $s$ of
$\l_2$, there is an experimental proposition $P$ on $S_2$ such
that $P$ is true if the state of $S_2$ is $r$ and false if the
state is $s$. Now, $P$ is a proposition concerning the compound
system $S$, and obviously for any atom $p$ of $\l_1$, $P$ is true
if the state of $S$ is $p\ot r$ and false if the state is $p\ot
s$. Therefore, as for propensity maps (see \cite{Gisin:1984} or
\cite{Piron:handbook}, \S 4.2), it is natural to assume that
\begin{enumerate}
\item[(A3)] $r\p_2 s\Rightarrow \omega(p_1\ot p_2,p\ot r\vee p\ot
s)=\omega(p_1\ot p_2,p\ot r)+\omega(p_1\ot p_2,p\ot s)$.
\end{enumerate}

From Axioms A1, A3 and Eq. (\ref{EquationPropensities}), we obtain
that for all atoms $p$ of $\l_1$ and $r,\,s$ of $\l_2$,
\begin{equation}\begin{split}r\p_2 s\ &\Rightarrow\
[\omega(p_1\ot p_2,p\ot r\vee p \ot s)=\omega(p_1\ot p_2,p\ot
(r\vee s)),\ \forall
p_1\ot p_2\in\sa(\l)]\\
&\Rightarrow p\ot r\vee p\ot s=p\ot (r\vee s)\,
.\end{split}\label{EquationPropensities2}\end{equation}

Now, for $i=1$ and $i=2$, let
\[f_i:\{V\in\pro(\h_i)\,;\,\mathrm{dim}(V)=2\}\rightarrow
2^{\sa(\pro(\h_i))}\, ,\]
such that for all $V$ in the domain of $f_i$, $f_i(V)$ is a
maximal set of mutually orthogonal atoms in $V^{\p_i}$. Moreover,
for any two atoms $p$ and $q$, define $A^{pq}_i:=\{p\}\cup
f_i(p\vee q)$. Suppose that $\mathrm{dim}(\h_i)\geq 4$. Let $p$
and $q$ be atoms. Then, $p\in A_i^{pq}$, $q\in A_i^{qp}$ and
$\abs{A_i^{pq}\cap A_i^{qp}}\geq 2$. Therefore, $\{A^{p q}_i\,;\,
p, q\in \sa(\pro(\h_i))\}$ forms a connected covering of
$\sa(\pro(\h_i))$. Moreover, from (\ref{EquationPropensities2}),
$\l$ is laterally connected.
\section{Automorphisms of $\mathsf{S}-$products}\label{SectionAutomorphims}

In this section, we show that automorphisms of
$\mathsf{S}-$products factor. We will use this result in the proof
of our main result (Theorem \ref{TheoremTheTheorem}).

\begin{theorem}\label{TheoremAutoFactor}
Let $\l_1,\,\l_2$ and $\l$ be complete atomistic lattices, with
$\l_1$ and $\l_2$ weakly connected and transitive. Let
$T_i\subseteq\aut(\l_i)$ acting transitively on $\sa(\l_i)$ with
$\id\in T_i$. Suppose that $\l$ is a
$\mathsf{S}_{_{T_1T_2}}-$product of $\l_1$ and $\l_2$. Then, for
any $u\in\aut(\l)$, there is a permutation $\xi$ of $\{1,2\}$, and
there are isomorphisms $u_i:\l_i\rightarrow \l_{\xi(i)}$, such
that for any atom, $u(p_1\ot p_2)=u_1(p_1)\ot u_2(p_2)$ if
$\xi=\id$ and $u(p_1\ot p_2)=u_2(p_2)\ot u_1(p_1)$ otherwise.
\end{theorem}

\begin{proof} The first three steps of the proof are similar to
those of the proof of Theorem 7.5 in \cite{Ischi:2004}. We denote
by $\{A_i^\gamma\}_{\gamma\in\sigma_i}$ the connected coverings of
Definition \ref{DefinitionLatterallyConnected}.

(1) {\bf Claim}\nobreak : For any atom $p=p_1\ot p_2$, we have
$u(p_1\ot 1)=u(p)_1\ot 1$ or $u(p_1\ot 1)=1\ot u(p)_2$. [{\it
Proof}\nobreak : Let $\gamma\in\sigma_2$ and $q,\, r\in
A_2^\gamma$. Since $\l$ is laterally connected, $p_1\ot q\vee
p_1\ot r$ contains a third atom, so does $u(p_1\ot q)\vee u(p_1\ot
r)$, for $u$ is join-preserving and injective. Thus, by Lemma
\ref{LemmaJoinins-Tensor} part 1, $u(p_1\ot q)$ and $u(p_1\ot r)$
differ only by one component. As a consequence, one of the
following cases holds: $u(p_1\oto A_2^\gamma)\subseteq u(p)_1\oto
1$, or $u(p_1\oto A_2^\gamma)\subseteq 1\oto u(p)_2$.

Define $f:\sigma_2\rightarrow \{1,2\}$ as $f(\gamma)=1$ if the
former case holds, and $f(\gamma)=2$ if the latter case holds.
Note that since $u$ is injective, if $\abs{A_2^\gamma\cap
A_2^\delta}\geq 2$, then $f(\gamma)=f(\delta)$.

Let $\gamma_0\in\sigma_2$ such that $p_2\in A_2^{\gamma_0}$. Then,
by the third hypothesis in Definition \ref{DefinitionConnected},
for all $q\in\sa(\l_2)$, there is $\gamma\in\sigma_2$ such that
$q\in A_2^\gamma$ and such that $f(\gamma)=f(\gamma_0)$. Hence,
for all $q\in\sa(\l_2)$, we have $u(p_1\ot
q)_{f(\gamma_0)}=u(p)_{f(\gamma_0)}$. As a consequence, either
$u(p_1\ot 1)\leq u(p)_1\ot 1$ or $u(p_1\ot 1)\leq 1\ot u(p)_2$.

Suppose for instance that the former case holds. Then $p_1\ot
1\leq u^{-1}(u(p)_1\ot 1)$, and since $u^{-1}$ is also
join-preserving and injective, $p_1\ot 1=u^{-1}(u(p)_1\ot 1)$.]

(2) From part 1, we can define
$g_2:\sa(\l_1)\oto\sa(\l_2)\rightarrow \{1,2\}$ as $g_2(p_1\ot
p_2)=2$ if $u(p_1\ot 1)=u(p)_1\ot 1$, and $g_2(p_1\ot p_2)=1$ if
$u(p_1\ot 1)=1\ot u(p)_2$. We define $g_1$ in an obvious similar
way ({\it i.e.} $g_1(p_1\ot p_2)=1$ if and only if $u(1\ot
p_2)=1\ot u(p)_2$). {\bf Claim}\nobreak : The maps $p\mapsto
g_i(p)$ are constant. [{\it Proof}\nobreak : Let $p=p_1\ot p_2$ be
an atom. Suppose that $g_2(p_1\ot p_2)=2$. Whence, $u(p_1\ot
1)=u(p)_1\ot 1$. Let $\gamma_0\in\sigma_1$ such that $p_1\in
A_1^{\gamma_0}$. Let $r\in A_1^{\gamma_0}$. Since $\l$ is
laterally connected, $p_1\ot p_2\vee r\ot p_2$ contains a third
atom, say $t\ot p_2$.

Let $v_2\in T_2$. By Axiom P4, $p_1\ot v_2(p_2)\vee r\ot v_2(p_2)$
contains $t\ot v_2(p_2)$. Therefore, since $\l_2$ is transitive,
we find that $t\ot 1\leq p_1\ot 1\vee r\ot 1$. Suppose now that
$u(r\ot 1)=1\ot(u(r\ot p_2))_2$. Then, by Lemma
\ref{LemmaJoinins-Tensor} part 2, we have
\[u(t\oto 1)\subseteq u(p)_1\oto1\cup 1\oto u(r\ot p_2)_2\, .\]
Therefore, $u(t\ot 1)=u(p_1\ot 1)$ or $u(t\ot 1)=u(r\ot 1)$, a
contradiction since $u$ is injective.

As a consequence, for all $r\in A_1^{\gamma_0}$, $u(r\ot 1)=u(r\ot
p_2)_1\ot 1$, hence $g_2(r\ot p_2)=g_2(p_1\ot p_2)$. Now, by the
third hypothesis in Definition \ref{DefinitionConnected}, we find
that $u(s\ot 1)=u(s\ot p_2)_1\ot 1$ for all $s\in\sa(\l_1)$.]

(3) Let $p=p_1\ot p_2$ be an atom. From part 2, we can define a
map $\xi:\{1,2\}\rightarrow\{1,2\}$ as $\xi(i):=g_i(p_1\ot p_2)$,
and $\xi$ does not depend on the choice of $p$. {\bf
Claim}\nobreak : The map $\xi$ is surjective. [{\it Proof}\nobreak
: Suppose for instance that $\xi(1)=1=\xi(2)$. Let $p=p_1\ot p_2$
and $q=q_1\ot q_2$ be atoms. Then
\[u(p)_2=u(1\ot p_2)_2=u(q_1\ot p_2)_2=u(q_1\ot 1)_2=u(q_1\ot
q_2)_2\, .\]
As a consequence, $u(1)\leq 1\ot u(p)_2$, a contradiction since
$u$ is surjective.]

(4) Let $p_1\ot p_2$ be an atom. For $i=1$ and $i=2$, define
$U_i:\sa(\l_i)\rightarrow\sa(\l_{\xi(i)})$ as $U_1(p):=u(p\ot
p_2)_{\xi(1)}$ and $U_2(q):=u(p_1\ot q)_{\xi(2)}$. {\bf
Claim}\nobreak : Those definitions do not depend on the the choice
of $p_1\ot p_2$. [{\it Proof}\nobreak : Suppose for instance that
$\xi=\mathsf{id}$. Then for any atom $r_2$ of $\l_2$, we have
\[u(p\ot p_2)_{\xi(1)}=u(p\ot 1)_{\xi(1)}=u(p\ot r_2)_{\xi(1)}\,
.]\]

Define $u_i:\l_i\rightarrow \l_{\xi(i)}$ as $u_i(a_i)=\vee
U_i(\sa(a_i))$. {\bf Claim}\nobreak : The map $u_i$ is an
isomorphism. [{\it Proof}\nobreak : Suppose for instance that
$\xi=\mathsf{id}$. Let $a\in\l_1$. Then, since $u$ and $h_1$ are
join-preserving, we find that
\[\begin{split}u(h_1(a))&=u(h_1(\vee\sa(a)))
=\vee\{u(h_1(r))\,;\,r\in\sa(a)\}=\vee\{u(r\ot 1)\,;\,r\in\sa(a)\}\\
&=\vee\{u(r\ot p_2)_1\ot 1\,;\,r\in\sa(a)\}=\vee\{h_1(u(r\ot
p_2)_1)\,;\,r\in\sa(a)\}\\
&=h_1(\vee\{u(r\ot
p_2)_1\,;\,r\in\sa(a)\})=h_1(\vee\{U_1(r)\,;\,r\in\sa(a)\})=h_1(u_1(a))\,
.\end{split}\]

As a consequence, since $h_1$ and $u$ are injective, so is $u_1$.
Let $\omega\subseteq\l_1$. Then, by the preceding formula, we find
that
\[\begin{split}h_1(u_1(\vee\omega))&=u(h_1(\vee\omega))=
\vee\{u(h_1(a))\,;\,a\in\omega\}\\
&=
\vee\{h_1(u_1(a))\,;\,a\in\omega\}=h_1(\vee\{u_1(a)\,;\,a\in\omega\})\,
.\end{split}\]
Whence, since $h_1$ is injective, $u_1$ preserves arbitrary joins.
Finally, since $U_1$ is surjective, so is $u_1$. As a consequence,
$u_1$ is a bijective map preserving arbitrary joins, hence an
isomorphism.]
\end{proof}
\section{Orthocomplemented $\mathsf{S}-$products}\label{SectionOrthocomplementation}

For our main result, we need some additional hypotheses on $\l_1$
and $\l_2$, which are true if $\l_1=\pro(\h_1)$ and
$\l_2=\pro(\h_2)$ with $\h_1$ and $\h_2$ complex Hilbert spaces.

\begin{definition}\label{DefinitionTransitive}
Let $\l$ be a complete atomistic lattice and let
$T\subseteq\aut(\l)$. We say that $\l$ is {\it $T-$strongly
transitive} if $\id\in T$, $T$ acts transitively on $\sa(\l)$, and
if
\begin{enumerate}
\item for all two $p\ne q\in\sa(\l)$, there is $u\in T$ such that
$u(p)=p$ and $u(q)\ne q$,
\item for all subset $\emptyset\ne A\subseteq\sa(\l)$, we
have:\hfill\break [$u(A)\cap A=u(A)$ or $\emptyset$, for all $u\in
T$] $\Rightarrow$ [$A=\sa(\l)$ or $A$ is a singleton].
\end{enumerate}
\end{definition}

\begin{lemma}\label{LemmaP(H)stronglyotransitive}
Let $\h$ be a complex Hilbert space. Then $\pro(\h)$ is
transitive. Moreover,
\begin{enumerate}
\item if $\mathrm{dim}(\h)\geq 3$, then $\pro(\h)$ is
$\unit(\h)-$strongly transitive,
\item if $\h=\mathbb{C}^2$ and if the second hypothesis in
Definition \ref{DefinitionTransitive} holds for some nonempty
subset $A\subseteq (\mathbb{C}^2-0)/\mathbb{C}$ and all
$u\in\unit(\mathbb{C}^2)$, then $A=\sa(\pro(\mathbb{C}^2))$, or A
is a singleton, or $A=\{p,p^\p\}$ with $p$ an atom of
$\pro(\mathbb{C}^2)$.
\end{enumerate}
\end{lemma}

\begin{proof} Obviously, $\mathsf{U}(\h)$ acts
transitively on $\sa(\pro(\h))=(\h-0)/\mathbb{C}$, and on each
coatom. Therefore, if $\mathrm{dim}(\h)\geq 3$, then the first
assumption in Definition \ref{DefinitionTransitive} holds.

We now check the second assumption of Definition
\ref{DefinitionTransitive}. Let $A\subseteq\sa(\pro(\h))$ be a
nonempty subset. Suppose that $A$ is not a singleton. Let $p,\,
q\in A$.

Assume first that $p\not\p q$. Define
\[\begin{split}G_p&:=\{u\in \mathsf{U}(\h)\,;\, u(p)=p\}\, ,\\
p\cdot q&:=\abs{ \langle P,Q\rangle}\, ,\end{split}\]
where $P\in p$, $Q\in q$ and $\Vert P\Vert=\Vert Q\Vert=1$.
Moreover, for $\omega\subseteq [0,1]$, define the cone
\[C_\omega(p):=\{r\in\sa(\pro(\h))\,;\, p\cdot r\in\omega\}\, .\]
Since $p\in u(A)\cap A$, for all $u\in G_p$, we have $C_{p\cdot
q}(p)\subseteq A$. Moreover, $C_{p\cdot r}(r)\subseteq A$, for all
$r\in C_{p\cdot q}(p)$. Therefore, we find that
$C_{[\lambda,1]}(p)\subseteq A$ where
\[\lambda=\max\{0,\cos(2\arccos(p\cdot q))\}\, ,\]
and furthermore that $\sa(\pro(\h))\subseteq A$.

Suppose now that $p\p q$. If $\mathrm{dim}(\h)\geq 3$, since ${\rm
dim}(p^\p)$ and $\mathrm{dim}(r^\p)$ are $\ge 2$, we find that
$\lambda=0$. Finally, if $\mathrm{dim}(\h)=2$ and $A=\{p,p^\p\}$,
then $u(A)\cap A= \emptyset$ or $u(A)$, for all $u\in
\unit(\mathbb{C}^2)$.
\end{proof}

\begin{lemma}\label{LemmaWeakTensorSTransitive}
Let $\l_1,\, \l_2$ and $\l$ be complete atomistic lattices, with
$\l_i$ $T_i-$strongly transitive for some
$T_i\subseteq\aut(\l_i)$. Suppose that $\l$ is a
$\mathsf{S}_{_{T_1T_2}}-$product of $\l_1$ and $\l_2$. Let
$R\subseteq \sa(\l)$ be nonempty, such that for all
$u\in\aut(\l)$, $u(R)\cap R=u(R)$ or $\emptyset$. Then one of the
following cases holds.
\begin{enumerate}
\item[(i)] $R=\sa(\l)$.
\item[(ii)] $R$ is a singleton.
\item[(iii)] $R=p\oto \sa(\l_2)$, for some $p\in\sa(\l_1)$.
\item[(iv)] $R=\sa(\l_1)\oto q$ for some $q\in\sa(\l_2)$.
\end{enumerate}
\end{lemma}

\begin{proof} (1) Let $p\in\sa(\l_1)$ and $q\in\sa(\l_2)$. Define
\[\begin{split}R(p)&:=\{s\in\sa(\l_2)\,;\,p\ot s\in R\}\, ,\\
R^{-1}(q)&:=\{r\in\sa(\l_1)\,;\,r\ot q\in R\}\, .\end{split}\]
{\bf Claim}\nobreak : If $R(p)\ne\emptyset$, then $R(p)=\sa(\l_2)$
or $R(p)=\{s\}$ for some $s\in\sa(\l_2)$. [{\it Proof}\nobreak :
Let $u_2\in T_2$ such that $u_2(R(p))\cap R(p)\ne\emptyset$. Then
$\id\ot u_2(R)\cap R\ne\emptyset$. Hence, by hypothesis, we have
$\id\ot u_2(R)\subseteq R$; therefore $u_2(R(p))\subseteq R(p)$.
As a consequence, the statement follows from the fact that $\l_2$
is $T_2-$strongly transitive.]

(2) Suppose that $p_1\ot p_2,\ q_1\ot q_2\in R$. Then, since
$\l_2$ is $T_2-$strongly transitive, there is $u_2\in T_2$ with
$u_2(p_2)=p_2$ and $u_2(q_2)\ne q_2$. As a consequence, $\id\ot
u_2(R)\cap R\ne\emptyset$, therefore, by hypothesis, $\id\ot
u_2(R)\subseteq R$. Hence, $\{q_2, u_2(q_2)\}\subseteq R(q_1)$.
Thus, by part 1, we have $R(q_1)=\sa(\l_2)$. In the same way, we
prove that $R(p_1)=\sa(\l_2)$. As a consequence,
$\abs{R^{-1}(y)}\geq 2$, for all $y\in\sa(\l_2)$. Therefore, by
part 1, $R^{-1}(y)=\sa(\l_1)$, for all $y\in\sa(\l_2)$, that is
$R=\sa(\l_1)\oto\sa(\l_2)=\sa(\l)$.\end{proof}

\begin{theorem}\label{TheoremTheTheorem}
Let $\l_1,\,\l_2$ and $\l$ be complete atomistic lattices, with
$\l_1$ and $\l_2$ coatomistic and weakly connected. Let
$(T_1,T_2)\subseteq\aut(\l_1)\t\aut(\l_2)$. Suppose that one of
the following cases holds.
\begin{enumerate}
\item[(i)] $\l_i$ is $T_i-$strongly transitive and $\l$ is a
$\mathsf{S}_{_{T_1T_2}}-$product of $\l_1$ and $\l_2$.
\item[(ii)] $\l_2$ is $T_2-$strongly transitive,
$\l_1=\pro(\mathbb{C}^2)$ and $\l$ is a
$\mathsf{S}_{_{\unit(\mathbb{C}^2)T_2}}-$product of $\l_1$ and
$\l_2$.
\item[(iii)] $\l_1$ is $T_1-$strongly transitive,
$\l_2=\pro(\mathbb{C}^2)$ and $\l$ is a
$\mathsf{S}_{_{T_1\unit(\mathbb{C}^2)}}-$product of $\l_1$ and
$\l_2$.
\item[(iv)] $\l_1=\pro(\mathbb{C}^2)=\l_2$ and $\l$ is a
$\mathsf{S}_{_{\unit(\mathbb{C}^2)\unit(\mathbb{C}^2)}}-$product
of $\l_1$ and $\l_2$.
\end{enumerate}
If $\l$ admits an orthocomplementation, then $\l$ is isomorphic to
$\l_1\aerts\l_2$, and $\l_1$ and $\l_2$ are orthocomplemented.
\end{theorem}

\begin{proof}We denote the dual of $\l_i$ (defined by the
converse order-relation) by $\l_i^*$. Hence $\sa(\l_i^*)$ stands
for the set of coatoms of $\l_i$.

Let $\p:\l\rightarrow\l$ be an orthocomplementation of $\l$. We
define a map $\Delta:\sa(\l_1^*)\t\sa(\l_2^*)\rightarrow \l$ as
\[\Delta(x_1,x_2)=(x_1\ot 1\vee 1\ot x_2)^\p=h_1(x_1)^\p\land
h_2(x_2)^\p\, .\]
Note that $0\ne \Delta(x_1,x_2)\ne 1$, for all
$(x_1,x_2)\in\sa(\l_1^*)\t\sa(\l_2^*)$. Moreover, $\Delta$ is
injective. We prove that the image of $\Delta$ is $\sa(\l)$, and
that $\Delta$ is a bijection. As a consequence, the map
$f:\l_1\aerts\l_2\rightarrow \l$ defined as $f(a)=\land
\{\Delta(x_1,x_2)^\p\,;\,
a\subseteq\sa(x_1)\t\sa(\l_2)\cup\sa(\l_1)\t\sa(x_2)\}$ is
obviously an isomorphism. As consequence, $\l_1\aerts\l_2$ is
orthocomplemented. Therefore, $\l_1$ and $\l_2$ are
orthocomplemented (see Proposition 8.3, \cite{Ischi:2004}). In the
sequel, we denote $\sa(\l_1^*)\t\sa(\l_2^*)$ by $\Xi$.

(1) {\bf Claim}\nobreak : For any two
$x=(x_1,x_2),\,y=(y_1,y_2)\in\Xi$, $\Delta(x)\land \Delta(y)=0$.
[{\it Proof}\nobreak : Suppose for instance that $x_1\ne y_1$, and
let $a\leq \Delta(x)\land \Delta(y)$. Then $x_1\ot 1\vee 1\ot
x_2\vee y_1\ot 1\vee 1\ot y_2\leq a^\p$. Therefore, since $h_1$
preserves joins and $1$, we have
\[1=h_1(1)=h_1(x_1\vee y_1)=h_1(x_1)\vee
h_1(y_1)\leq a^\p\, ;\]
whence $a^\p=1$, that is $a=0$.]

(2) {\bf Claim}\nobreak : For all $x=(x_1,x_2)\in\Xi$ and all
$u\in{\rm Aut}(\l)$, there is $y\in\Xi$ such that
$u(\Delta(x))=\Delta(y)$. [{\it Proof}\nobreak : First note that
$u^\p:\l\rightarrow\l$ defined as $u^\p(a)=u(a^\p)^\p$ is an
automorphism of $\l$. By Theorem \ref{TheoremAutoFactor}, there
are two isomorphisms $v_1$ and $v_2$ and a permutation $\xi$ such
that for any atom, $u^\p(p_1\ot p_2)_{\xi(i)} =v_i(p_i)$. Suppose
for instance that $\xi=\mathsf{id}$. Then,
\[u(\Delta(x))=(u^\p(\Delta(x)^\p))^\p=(u^\p(x_1\ot 1
\vee 1\ot x_2))^\p= (v_1(x_1)\ot 1\vee 1\ot v_2(x_2))^\p\, ,\]
hence $u(\Delta(x))= \Delta(v_1(x_1),v_2(x_2))$.]

(3) {\bf Claim}\nobreak : $\cup\{\sa(\Delta(x));\
x\in\Xi\}=\sa(\l)$. [{\it Proof}\nobreak : Let $x\in\Xi$. By
Axioms P5 and P4, $\l$ is transitive. As a consequence, for any
atom $r$ of $\l$, there is an automorphism $u$ such that $r\leq
u(\Delta(x))$, hence by part 2, there is $y\in\Xi$ such that
$r\leq \Delta(y)$.]

(4) Consider assumption (i). {\bf Claim}\nobreak : For all
$x\in\Xi$, $\Delta(x)$ is an atom. [{\it Proof}\nobreak : Let
$x^0=(x^0_1,x^0_2)\in\Xi$. From part 2 and 1, $u(\Delta(x^0))\land
\Delta(x^0)=u(\Delta(x^0))$ or $0$, for all $u\in \aut(\l)$.
Therefore, by Lemma \ref{LemmaWeakTensorSTransitive}, either
$\Delta(x^0)$ is an atom, or $\Delta(x^0)=r\ot 1$ for some
$r\in\sa(\l_1)$, or $\Delta(x^0)=1\ot s$ for some $s\in\sa(\l_2)$.

Suppose for instance that $\Delta(x^0)=r\ot 1$. Then, since $\l_1$
is transitive, by part 2, for all $s\in\sa(\l_1)$, there is
$y\in\Xi$ such that $s\ot 1=\Delta(y)$, hence by part 1, for all
$z\in\Xi$, there is $s\in\sa(\l_1)$ such that $\Delta(z)=s\ot 1$.
Therefore, there is a bijection $f:\Xi\rightarrow\sa(\l_1)$ such
that $\Delta(x)=f(x)\ot 1$, for all $x\in\Xi$.

Let $t\leq \Delta(x^0)$ be an atom. By Axiom P2, since $\l_1$ and
$\l_2$ are coatomistic, we have that $\land\{\Delta(x)^\p;\ t\leq
\Delta(x)^\p \}=t$; whence
\[\begin{split}x^0_1\ot 1\vee 1\ot x^0_2=\Delta(x^0)^\p\leq t^\p&
=\vee\{\Delta(x);\ t\leq \Delta(x)^\p\}\\
&=\vee\{f(x)\ot 1;\ t\leq \Delta(x)^\p\}= a\ot 1 \, ,\end{split}\]
(with $a=\vee\{f(x)\,;\,t\leq \Delta(x)^\p\}$) a contradiction. As
a consequence, $\Delta(x^0)$ is an atom.]

(5) Consider now assumption (ii). Suppose that none of the cases
treated in part 4 holds for $\Delta(x_0)$. Then by the same
argument as in Lemma \ref{LemmaWeakTensorSTransitive}, we find
that $\sa(\Delta(x_0))=\{r,r^{\p_1}\}\oto 1$ or
$\{r,r^{\p_1}\}\oto s$ with $r$ and $s$ atoms. Both cases can be
excluded from the last requirement in Definition
\ref{DefinitionLatterallyConnected}.

(6) Finally, consider assumption (iv). The last case we have to
exclude is $\sa(\Delta(x_0))=\{r\ot s,r^{\p_1}\ot s^{\p_2}\}$.

(6.1) Let $G:=\unit(\mathbb{C}^2)\t \unit(\mathbb{C}^2)$.  {\bf
Claim}\nobreak : For all $p,\, q\in\sa(\l)$, there is
$(u_1,u_2)\in G$ such that $u_1\ot u_2(p^\p)=q^\p$. [{\it
Proof}\nobreak : Let $g$ be the action of $G$ on $\sa(\l)$ defined
as $g(u_1,u_2)(p)=(u_1\ot u_2)^\p(p)$ (see part 2). Let $p$ and
$q$ be atoms of $\l$. Then there is $(u_1,u_2)\in G$ such that
$u_1\ot u_2(p)=q$, thus $(u_1\ot u_2)^\p(p^\p)=q^\p$. Hence, $G$
acts transitively on the set of coatoms of $\l$.

{\bf Claim}\nobreak : $G$ acts transitively on the set of coatoms
of $\l_1\aerts\l_2$. [{\it Proof}: Let $x\ne y\in\Xi$. By Axiom
P2, there are atoms $r$ and $s$ such that $\Delta(x)^\p\leq r^\p$
and $\Delta(y)^\p\leq s^\p$. By what precedes, there is
$(u_1,u_2)\in G$ such that $(u_1\ot u_2)^\p(r^\p)=s^\p$, hence
$(u_1\ot u_2)^\p(\Delta(x)^\p)\leq s^\p$. Since by Theorem
\ref{TheoremAutoFactor}, $(u_1\ot u_2)^\p$ factors, from part 1,
$(u_1\ot u_2)^\p(\Delta(x)^\p)=\Delta(y)^\p$.]

As a consequence, since $\l_1$ and $\l_2$ are of length 2, the
action of $G$ on $\sa(\l)$ is transitive. Therefore, for all $p,\,
q\in\sa(\l)$, there is $(u_1,u_2)\in G$ such that $u_1\ot
u_2(p^\p)=q^\p$.]

(6.2) Let $p\in \sa(\Delta(x_0))$, then $\Delta(x_0)^\p\leq p^\p$.
From part 6.1, for any coatom $q^\p\geq \Delta(x_0)^\p$, there is
$(u_1,u_2)\in G$ such that $u_1\ot u_2(p^\p)=q^\p$. Therefore,
\[ \{q^\p\,;\,q^\p\geq \Delta(x_0)^\p\}=\{u_1\ot u_2(p^\p)
\,;\,(u_1,u_2)\in
G_{x_0}\}\, ,\]
where $G_{x_0}:=\{(u_1,u_2)\in G\, ;\,
(u_1({x_0}_1),u_2({x_0}_2)=x_0\}$. Whence,
\[\sa(\Delta(x_0))=\{u_1\ot u_2(p^\p)^\p\,;\, (u_1,u_2)\in
G_{x_0}\}\, .\]

If $\sa(p^\p)$ is invariant under the action of $G_{x_0}$ ({\it
i.e.} $u_1\ot u_2(p^\p)=p^\p$, for all $(u_1,u_2)\in G_{x_0}$),
then $\Delta(x_0)=p\in\sa(\l)$.

Otherwise, write $\sa(p^\p)$ as
\[\sa(p^\p)=\bigcup_{\alpha\beta} c_\alpha({x_0}_1)\oto
c_\beta({x_0}_2)\, ,\]
where $c_\cdot({x_0}_i)$ is included in the cone
$C_\cdot({x_0}_i)$ around ${x_0}_i$ (see the proof of Lemma
\ref{LemmaP(H)stronglyotransitive}). Now, if $\sa(p^\p)$ is not
invariant under the action of $G_{x_0}$, then there is
$c_\cdot({x_0}_i)$ which is a proper subset of $C_\cdot({x_0}_i)$.
Therefore, obviously $\Delta(x_0)$ contains more than two atoms.
\end{proof}

\begin{remark}\label{Remarkuorthoautomorphism}
Note that  if $\l_1$ and $\l_2$ are orthocomplemented and $h_1$
and $h_2$ preserve also the orthocomplementation, then for any
atom, we have
\[(p_1\ot p_2)^\p=(h_1(p_1)\land h_2(p_2))^\p=h_1(p_1)^\p
\vee h_2(p_2)^\p=h_1(p_1^{\p_1})\vee h_2(p_2^{\p_2})\, ,\]
so that the proof is trivial. On the other hand, if we ask the $u$
of Axiom P4 to be an ortho-automorphism, then for all $x\in\Xi$,
we have $u_1\ot u_2(\Delta(x))=\Delta(u_1(x_1),u_2(x_2))$, so that
part 2 of the proof becomes trivial, and the proof does not
require Theorem \ref{TheoremAutoFactor}.
\end{remark}
\bibliographystyle{abbrv}
\bibliography{Ischi-Final}
\end{document}